\documentclass
[twocolumn,showpacs,preprintnumbers,amsmath,amssymb,aps,prl,superscriptaddress,floatfix,reprint]
{revtex4}
\usepackage{graphicx}
\usepackage{dcolumn}
\usepackage{bm}
\usepackage{here}    
\usepackage{color}
\usepackage{ulem}

\begin{document}


\title{Lifetime of the arrow of time inherent in chaotic eigenstates\\
:case of coupled kicked rotors}

\author{Fumihiro Matsui}
\email{fumihiro-matsui[at]xnea.net}
\affiliation{Demartment of Physics, College of Science and Engineering, Ritsumeikan University,
Noji-higashi 1-1-1, Kusatsu 525-8577, Japan}

\author{Hiroaki S. Yamada}
\affiliation{Yamada Physics Research Laboratory, Aoyama 5-7-14-205, Niigata 950-2002, Japan}

\author{Kensuke S. Ikeda}
\email{ahoo[at]ike-dyn.ritsumei.ac.jp}
\affiliation{College of Science and Engineering, Ritsumeikan University,
Noji-higashi 1-1-1, Kusatsu 525-8577, Japan}

\newcommand{\vc}[1]{\mbox{\boldmath $#1$}}
\newcommand{\fracd}[2]{\frac{\displaystyle #1}{\displaystyle #2}}
\newcommand{\red}[1]{\textcolor{red}{#1}}
\newcommand{\blue}[1]{\textcolor{blue}{#1}}
\newcommand{\green}[1]{\textcolor{green}{#1}}
\newcommand{\del}{\partial}

\date{\today}
\begin{abstract}
A linear oscillator very weakly coupled with the object quantum system
is proposed as a detector measuring the lifetime of irreversibility 
exhibited by the system, 
and classically chaotic coupled kicked rotors are examined as ideal examples. 
The lifetime increases drastically in close correlation with the 
enhancement of entanglement entropy(EE) between the kicked rotors.
In the transition regime to the full entanglement, the EE of individual 
eigenstates fluctuates anomalously, and the lifetime also fluctuates in 
correlation with the EE. In the fully entangled regime the 
fluctuation disappear, but the lifetime is not yet unique but 
increases in proportion to the number of superposed eigenstates 
and is proportional to the square of Hilbert space dimension in 
the full superposition.
\end{abstract}

\maketitle


\def\tr#1{\mathord{\mathopen{{\vphantom{#1}}^t}#1}} 

\def\ni{\noindent}
\def\nn{\nonumber}
\def\bH{\begin{Huge}}
\def\eH{\end{Huge}}
\def\bL{\begin{Large}}
\def\eL{\end{Large}}
\def\bl{\begin{large}}
\def\el{\end{large}}
\def\beq{\begin{eqnarray}}
\def\eeq{\end{eqnarray}}

\def\eps{\epsilon}
\def\th{\theta}
\def\del{\delta}
\def\omg{\omega}

\def\e{{\rm e}}
\def\exp{{\rm exp}}
\def\arg{{\rm arg}}
\def\Im{{\rm Im}}
\def\Re{{\rm Re}}

\def\sup{\supset}
\def\sub{\subset}
\def\a{\cap}
\def\u{\cup}
\def\bks{\backslash}

\def\ovl{\overline}
\def\unl{\underline}

\def\rar{\rightarrow}
\def\Rar{\Rightarrow}
\def\lar{\leftarrow}
\def\Lar{\Leftarrow}
\def\bar{\leftrightarrow}
\def\Bar{\Leftrightarrow}

\def\pr{\partial}

\def\Bstar{\bL $\star$ \eL}

\def\>{\rangle}
\def\<{\langle}

\def\H{\hat{H}}
\def\U{\hat{U}}
\def\Tr{\rm Tr}
\def\v{\hat{v}}
\def\J{\hat{J}}
\def\p{\hat{p}}
\def\q{\hat{q}}
\def\corr{C{\rm r}}

\def\vq{\hat{\vc{q}}}
\def\vp{\hat{\vc{p}}}
\def\vQ{\vc{Q}}
\def\vP{\vc{P}}

\def\T{{\cal T}}
\def\U{\hat{U}}
\def\tlife{\tau_L}
\def\avrtlife{\langle \! \langle \tau_L \rangle \! \rangle}
\def\dimN{N_{\rm dim}}

The classical chaotic system can be the simplest origin of irreversibility and 
the arrow of time \cite{prigogine}. Its quantum counterpart can also be the minimal unit 
exhibiting quantum irreversibility. Indeed, normal diffusion \cite{casati,fishman}, 
energy dissipation \cite{ikedadissip}, energy spreading \cite{cohen} and many apparently 
irreversible phenomena can be realized in classically chaotic quantum systems with a 
small number of degrees of freedom. 
Appearance of irreversibility in closed quantum system is also a severe 
problem limiting the performance of quantum computation algorithm \cite{quantumcomputation}.
If the chaotic region is bounded in the phase space and thus 
the effective dimension $\dimN$ of the subspace of Hilbert space relevant 
for the chaotic region is 
finite, the irreversibility can not be sustained in time infinitely. 
Even though $\dimN$ is infinite and the system is unbounded, the persistent 
coherence and the localization nature inherent in quantum system prevents 
the system from the complete manifestation of the irreversibility \cite{casati,fishman}. 
However, the multidimensional unbounded quantum chaos systems can recover classical
chaotic nature \cite{quantumclassical} and recover a complete irreversible 
behavior via the Anderson-transition-like mechanism 
under appropriate condition \cite{casatianderson,delande}.

Irreversibility in quantum system has been explored by the time-reversal
characteristics \cite{timerevexp,timerevexp-ballentine,benenti}. 
In particular, it has been extensively investigated 
by many investigators in the context of fidelity \cite{fidelity1,fidelity2}. 
The time scale on which quantum chaos can show the exponential sensitivity is 
quite short and is up to the Ehrenfest time proportional to $\log(\dimN)$  at 
the most \cite{bermanehrenfest}, but the time scale on which the irreversibility is 
maintained is much longer and is said to be as long as the Heisenberg time, 
which is proportional to $\dimN$ \cite{fidelity2}. 

It seems to be very difficult to give a general definition for the lifetime in 
which the quantum irreversibility is sustained, but it would be a very important 
parameter characterizing quantitively the irreversibility,
 i.e., the time of arrow self-organized in closed quantum systems.
The main purpose of the present article is to propose a general method for observing
the irreversibility inherent in quantum systems and its associated quantum states
and to measure its lifetime, taking typical quantum chaos systems as examples.
%
An important result is that the lifetime measured by the proposed method 
is in general much longer than the conventional Heisenberg time.

We have the fidelity as a powerful tool extracting the irreversible characteristics
of quantum system \cite{fidelity2}. 
However, the fidelity is not convenient for the purpose 
of observing irreversible characteristics with adequate numerical accuracy
over an extremely long time scale without disturbing the examined system.

We introduce a linear oscillator which converts the quantum motion of the 
object system to a Brownian motion in the infinitely extended linear oscillator's 
action space.
Let the Hamiltonian of the examined quantum system ``S'' be $\H(\vp,\vq,t)$
~($\vq$ and $\vp$ are coordinate and momentum vector operator, respectively)
and it is very weakly coupled with the linear oscillator ``L'', which is 
represented by angle-action canonical pair operators $\hat{\theta}$ and 
$\J=-i\hbar d/d\theta$ with the Hamiltonian $\omega \J$ of the frequency $\omega$. 
The total Hamiltonian reads 
\begin{eqnarray}
\label{hamil}
   \H_{tot}(t) = \H(\vp,\vq,t) + \eta \v(\vp,\vq)g(\hat{\theta}) + \omega \J, 
\end{eqnarray}
where $\v$ is a Hermitian operator, and $g(\hat{\theta})$ is $2\pi$-periodic
function with null average \cite{shepe83}. 
By monitoring the fictious diffusion in the $J$-space we can 
measure the lifetime of irreversibility without disturbing the object by
reducing $\eta$ as small as possible in accordance with the time of observation.

$\J$ has the eigenvalue $J=j\hbar~(j\in {\bf Z})$ for the eigenstate $\<\theta|J\>\propto\e^{-iJ\theta/\hbar}$
because of the $2\pi$-periodicity in $\theta$-space. 
We further impose the action periodic boundary condition identifying $J=L\hbar$ with $J=-L\hbar$, 
which quantize $\theta$ as $\theta_k=2\pi k/(2L)~~(k\in {\bf Z})$, where $-L<k\leq L$.

We are interested in the lifetime of irreversibility realized by classically chaotic quantum system
defined in a bounded phase space. To be concrete, we hereafter confine ourselves to the particular 
case that the system $\H(\vp,\vq,t)$ consists of 
the coupled kicked rotors (CKR) 
which exhibits ideal chaos.
More general properties of the model will be presented in a forthcoming publication \cite{matsuifull}. 
We consider here the coupled kicked rotors (CKR) $H(\vp,\vq,t)=\sum_{i=1}^2[\p_i^2/2+V(\q_i)\Delta(t)]+
\eps V_{12}(\q_1,\q_2)\Delta(t)$,  as a sample system S, where $\Delta(t)=\sum_{\ell}\delta(t-\ell T)$, 
because the lifetime of irreversibility will vary with the development of entanglement 
between the two constituent chaotic rotors controlled by the coupling strength $\eps$ 
\cite{couplerotorentangle}.

\def\Sm1{S_m^{(1)}}
\def\rhom1{\rho_m^{(1)}}
KR is observed at the integer multiple of the fundamental period $T$ as $t=\tau T+0$ where
$\tau \in {\bf Z}$. Then the one step evolution of CKR from $t=\tau T+0$
to $t+\tau$ is described by the unitary operator $\U=\e^{-i[\p_1^2+\p_2^2]/2\hbar}\e^{-i[V(\q_1)
+V(\q_2)+\eps V_{12}(\q_1,\q_2)]/\hbar}$. Each KR, say KR1 and KR2, is defined in the 
bounded phase space $(q_i,p_i)\in [0,2\pi]\times[0,2\pi]$ 
and 
$\hbar=2\pi/N_i$, where $N_1=N_2$ is a positive integer, and the dimension of
the Hilbert space for the CKR is $N=N_1N_2=N_1^2$, 
which is an important parameter. 
We take mainly the Arnold's cat map $V(\q)=-K\q^2/2$ (and also the standard map $V(\q)=K\cos \q$, if necessary)
with the interaction $V_{12}(\q_1,\q_2)=\cos(\q_1-\q_2)$. 
Hereafter, we couple L with S via only the the first KR (KR1) as 
 $\hat{v}(\p,\q)=\hat{v}(\p_1)=\sin(\p_1)$ 
so that the entanglement between the KRs may sensitively be reflected in the dynamics of L, 
and take the initial states of L and S as $|J=0\>$ and $|\Psi_0\>$, respectively. 
We reduce the coupling $\eta$ weak enough to eliminate the backaction from L to S, 
then the mean square displacement (MSD) of action of L reads  
\begin{eqnarray}
\label{J2corr}
  \<\J^2(\tau)\>= \sum_{s=0}^{\tau-1} D(s),~ D(\tau)=D_0\sum_{s=-\tau}^\tau \corr_\tau(s)\cos(\omega Ts)
\end{eqnarray}
where  $\<...\>$ means the expectation value with the initial condition $|\Psi_0\> \otimes |J=0\>$,  
and $D_0 =2\eta^2\sin^2(\omega T/2)/\omega^2$. Note that $\<\J(\tau)\>=0$
thanks to the initial state $|J=0\>$. $\corr_\tau(s)=(\<\Psi_0|\v_{\tau} \v_{\tau-s}|\Psi_0\> + c.c.)/2$, 
where $\hat{v}_{\tau}=\U^{-\tau}\hat{v}\U^\tau$ and $g(\hat{\theta})=\cos(\hat{\theta}-\omega T/2)$
, 
is the autocorrelation function satisfying $\corr_\tau(-s)=\corr_\tau(s)$ labelled by
$\tau (\in {\bf Z})$.

For classical KR having the ideally chaotic property such as K- and C-systems, 
the correlation function decays exponentially, and $D(t)$ coincides with the classical 
diffusion constant $D_{cl}$. However, in finitely bounded quantum KR, $D(t)$ 
tentatively approaches to $D_{cl}$ but finally it goes down to 0 on average, 
which means $\<\J^2\>$ saturates at a finite value, say $J_\infty^2$
as $\tau \to \infty$. 

\begin{figure}[htbp]
\begin{center}
\includegraphics[height=5cm]
{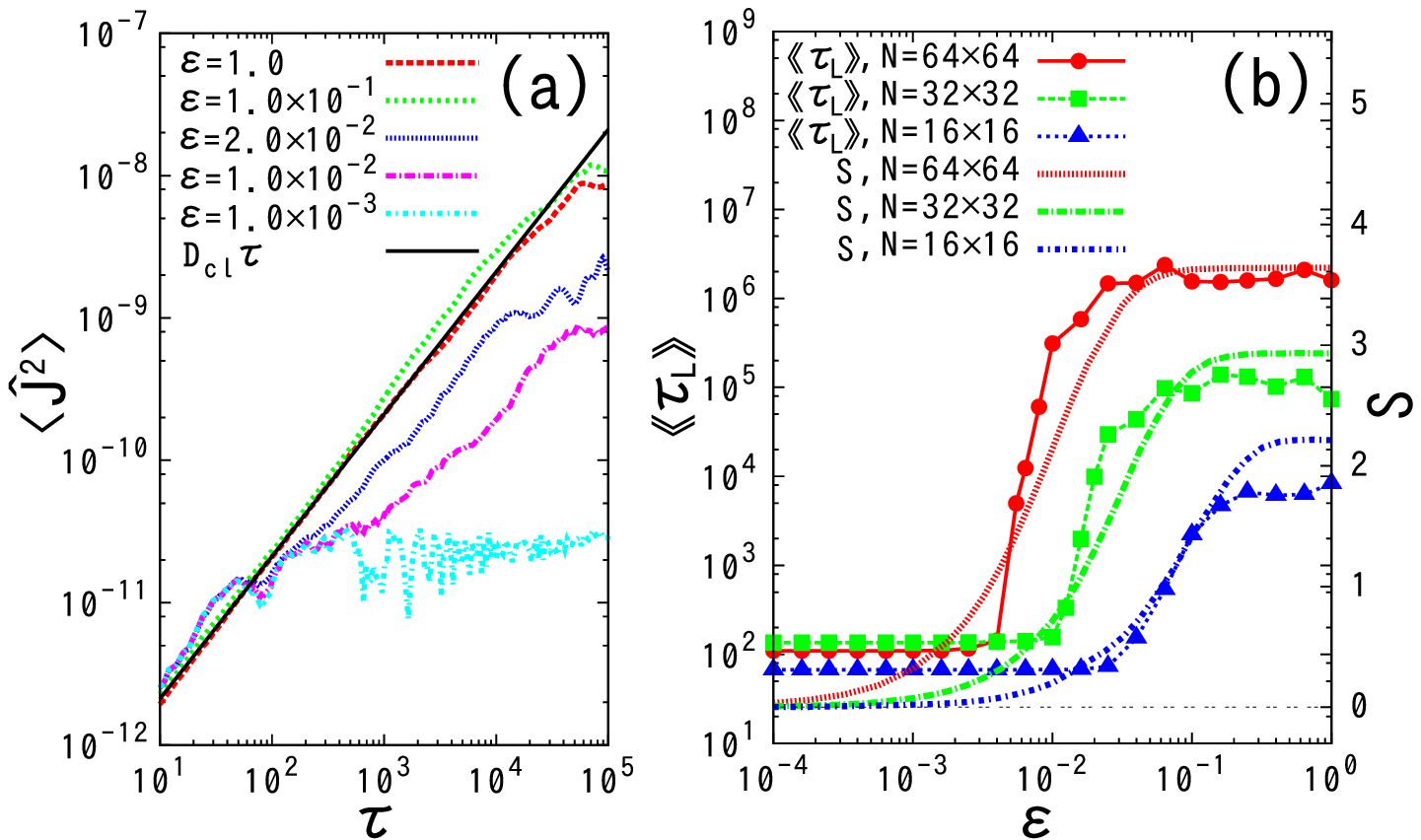}
\caption{\label{Fig1}(a)$\<\J^2\>$ vs $\tau$ for various values of $\eps$, 
where $N=32\times32$ and K=10.
(b) The EE $S$ vs $\eps$(the right vertical axis) 
and $\avrtlife$ vs $\eps$(the  left vertical axis) 
for $N=16\times16,~32\times32$, and $64\times64$, respectively}
\end{center}
\end{figure}
We define the lifetime $\tlife$ of the irreversibility that is represented as the stationary 
diffusion of L as follows: decide the diffusion exponent $\alpha(\tau)$ defined for $s$ in
an appropriate interval $[\tau-\Delta\tau/2,\tau+\Delta\tau/2]$ around $\tau$ such that 
$\<\J^2(s)\> \propto s^{\alpha(\tau)}$, we define $\tlife$ as the first step that deviate appreciably 
from $1$ such that $|\alpha(\tlife)-1|>r$, where we choose $r=0.5$ typically.

The lifetime thus defined in general varies sensitively with the choice of $|\Psi_0\>$.
In order to eliminate such accidental fluctuations, we add
small term such as $\xi_{iR}\cos(\q_i-q_{iR})~(\xi_{iR}\sim O(\hbar))$ to 
the potential $V(\q_i)$, and take the average of $\tlife$ with respect to 
the ensemble of the potential parameter $q_{iR}~(i=1,2)$. 
We refer it hereafter as the average lifetime $\avrtlife$.

Figure \ref{Fig1}(a) shows $\<\J^2\>$ vs $\tau$ at various $\eps$ for the coupled Arnold's cats. 
They all saturate at a finite level, but there seems to be a critical value 
$\eps_c\sim \hbar^2$ beyond which the saturation level is drastically enhanced. 
We are interested in how this enhancement is related to the development of entanglement between two KRs. 
To measure the entanglement degree we introduce the von-Neumann entropy as 
the entanglement entropy (EE). 
Let the quasi-eigenstate of $\U$ be $|m\>$, that is, $\U|m\>=e^{-i\gamma_m}|m\>$, where $\gamma_m$ is 
the eigenangle, then the EE is given by $\Sm1=-\Tr(\rhom1\log\rhom1)$ for the reduced density operator
$\rhom1={\rm Tr}_2 |m\>\<m|$ traced over the second cat. 

As shown in Fig.\ref{Fig1}(b), the average entropy $S=\sum_{m=1}^N\Sm1/N$ 
increases around a threshold value $\eps_c$ from 0 to a saturated value.
This fact means the two systems are entangled beyond $\eps=\eps_c$ to form a fully entangled 
two-degrees of freedom system \cite{couplerotorentangle}. 
A quite important fact is that the average lifetime increases 
in agreement with the increase of EE, which means that the lifetime of irreversibility grows in 
a strong correlation with the entanglement {\it on average}.

\begin{figure}[htbp]
\begin{center}
\includegraphics[height=5cm]
{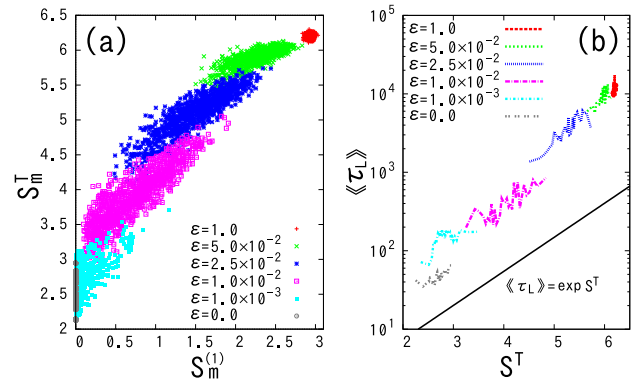}
\caption{\label{Fig2} (a) Plots of EE vs the transition entropy $(S_m^{(1)},S_m^T)$ 
for every eigenstate $|m\>$ of 
the coupled cat with $N=32\times32,~K=10$. 
The broadly scattered $S_m^{(1)}$ in the transition region (typically $\eps=0.01$ and $0.025$) 
indicate the anomalous fluctuation. Evidently $S_m^{(1)}$ is correlated with $S_m^T$. 
(b) The average lifetime as a function of the transition entropy $S^T$.}
\end{center}
\end{figure}

Unlike the unbounded CKR, the above transition is not accompanied by 
critical phenomena \cite{delande}, but the entropy of individual eigenstate shows an anomalous 
fluctuation in the transition regime, as shown in Fig.\ref{Fig2}(a). 
Widely distributed EE means a rich variety of the degree of entanglement in eigenstates, 
which leads us to expect the lifetime $\tlife$ of each eigenstate also fluctuates in a 
correlation with EE. However, the accidental fluctuation of $\tlife$ is so large that 
it is very difficult to observe a direct correlation between $\Sm1$ and $\tlife$. 

Instead, we can show that the lifetime and EE is correlated indirectly. 
Since $\v(\p_1)$ is free of KR2, the connection between eigenstates represented by 
$\<m|\v|n\>$ is enhanced only if KR1 and KR2 become entangled. 
The transition entropy  $S^T_m$ representing the variety of transition by the
perturbation $\v$ can be defined as $S^T_m=-\sum_{n}t_{mn}\log t_{mn}$, where 
$t_{mn}=|\<m|\hat{v}|n\>|^2/\sum_{n'} |\<m|\hat{v}|n'\>|^2$. It would have a strong correlation 
with the EE. In Fig.\ref{Fig2}(a) we the plot $(\Sm1,S^T_m)$ for every $|m\>$
, where the wide spread of $\Sm1$ in the transition region of $\eps$ indicate 
the anomalous fluctuation stated above. It is evident that $S_{m}^{(1)}$ has a 
strong correlation with $S_m^T$.

On the other hand, the entropy $S^T_m$ measures the number 
$B_m$ of the bonds connecting $|m\>$ with other states $|n\>$ by the relation 
%
$S_m^T\sim\log(B_m)$ or $B_m\sim\e^{S_m^T}$.
%
%
Here we suppose that the Heisenberg time is the time scale representing the lifetime,
then it is proportional to the number of the relevant quantum states contributing 
to the irreversible behavior, which is nothing more than $B_m$ in the present case.
We are thus lead to examine the correlation between lifetime and transition entropy
$S_m^T\sim\log B_m$.
%
Sorting $|m\>$ in the order of $S^T_m$ and superposing the same 
number of $|m\>$ with $S^T_m$ around a given $S^T$, 
we construct $|\Psi_0\>$ and measure the lifetime $\tlife$ of the state. 
We further take its ensemble average and show $\avrtlife$ as a 
function of $S^T$ in Fig.\ref{Fig2}(b). 
Evidently $\log(\avrtlife)$ increases linearly  with $S^T$.

Figure \ref{Fig2}(a) and (b) tell that $\avrtlife$ is correlated with entanglement 
entropy  $\Sm1$ via the transition entropy. 
This fact implies a quite interesting fact that at 
the {\it birth of the irreversibility} the degree of entanglement fluctuates
anomalously, which is reflected in the broad variety of lifetime of irreversibility
realized by each eigenstate.


\begin{figure}[htbp]
\begin{center}
\includegraphics[height=5cm]
{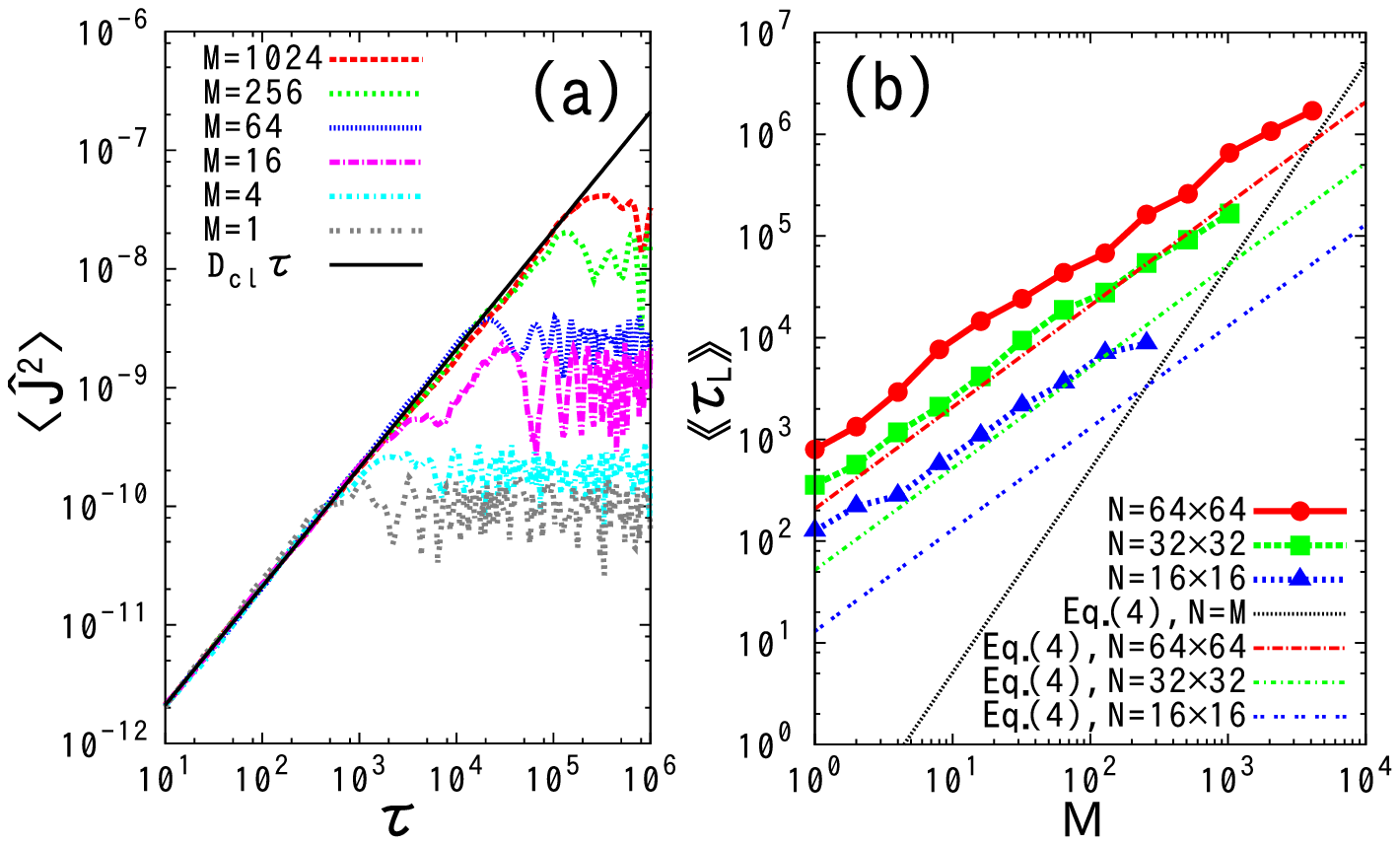}
\caption{\label{Fig3}(a) $\<\J^2\>$ vs $\tau$ for various $M$ in the fully entangled regime
$\eps=1.0$,  and $K=10$. 
(b) $\avrtlife$ vs $M$ for $N=16\times16,~32\times32,~64\times64$ denoted by 
some symbols.
Equation (\ref{tlife}) are indicated by the broken lines in the regime.}
\end{center}
\end{figure}
Beyond the threshold value $\eps (\gg \eps_c)$, EE saturates, and $B_m$ becomes the largest
value $N$, and the lifetime $\tlife$ becomes the longest one. 
Here the classicalization of dynamics 
of CKR is completed \cite{quantumclassical}, which drives the classical diffusion of L for $\tau \leq \tlife$.
However, a nontrivial problem arises here. See Fig.\ref{Fig3}: the time evolution pattern 
of MSD markedly changes with the number $M$ of superposing eigenstates taken as 
the initial state $|\Psi_0\>=\sum_{m=1}^M C_{m}|m\>$ 
with the typical weight $|C_m|^2 \sim O(1/M)$.
The lifetime at which the diffusion 
terminates increases remarkably with $M$ although the variation is not systematic. 
To make sure the above observation we show in Fig.\ref{Fig3}(b) how the average 
lifetime $\avrtlife$ varies with $M$. The lifetime $\avrtlife$ 
is proportional to $N$ if $M=1$, and it increases in proportion to $M$, which means
$\avrtlife \propto N^2$ in the fully mixed limit $M=N$. Such a behavior is
observed irrespective of the natural frequency $\omega$ as long as $\omega\neq 0$.

The above results are hardly expected, and we hereafter consider the reason closely. 
To this end we evaluate the saturation level $J_\infty^2$ of MSD by averaging Eq.(\ref{J2corr}) 
over $\tau$. A straightforward calculation yields
\begin{eqnarray}
\label{satu}
\frac{J_\infty^2}{D_0}=\sum_{m=1}^M|C_m|^2\sum_{n=1}^{N}
\frac{|\<m|\hat{v}|n\>|^2}{4}[\frac{1}{\sin^{2}(\frac{\delta_{mn}^+}{2})}
+\frac{1}{\sin^2(\frac{\delta_{mn}^-}{2})}], 
\end{eqnarray}
where $\delta_{mn}^\pm=\gamma_m-\gamma_n\pm \omega$.
Here the interference terms between 
different $m$s are neglected (the diagonal approximation) because 
their contribution is negligibly small in the regime we are concerned. 

First we consider the very particular case of $\omg=0$ in order to 
compare with general case discussed later. As is well-known in the study 
of fidelity, if $\v$ has diagonal components, it leads to a ballistic 
diffusion of L beyond the Heisenberg time \cite{fidelity2}. Therefore, we 
consider the case of null diagonal component. Then the summation over $n$ 
in the RHS of Eq.(\ref{satu}) is dominated by the component with minimal 
$\sin^2(\delta_{mn}^\pm/2) \sim |\gamma_m-\gamma_n|^2/4$. 
It is nothing more than the nearest 
neighbouring eigenangle distance following the Wigner distribution which have 
{\it a definite peak} around $2\pi/N$. Thus the magnitude of each term is dominated by 
$\sin(\delta_{mn}^\pm/2)\sim (\pi/N)^2$ irrespective of $m$, while by using the 
expression for the autocorrelation function at $s=0$  
it follows that $\corr(0)=\sum_m |C_m|^2 \sum_n|\<m|\hat{v}|n\>|^2$ (the subscript $\tau$ of 
$\corr$ can now be omitted) the representative value $\<|v^2|\>$ of 
the matrix element $|\<m|\hat{v}|n\>|^2$ is evaluated as $\<|v^2|\>=\corr(0)/N$, 
because the full mixing in this regime makes the magnitude of matrix elements $|\<m|\hat{v}|n\>|^2$ 
almost equal. As a result we evaluate the saturation level as $J_\infty^2\sim D_0C(0)N/2\pi^2$. 
Lifetime is the time at which the classical diffusion $D_{cl}\tau$ terminates 
at the saturation level, namely $D_{cl}\tlife=J_\infty^2$. It leads to 
$\tlife = \frac{D_0\corr(0)N}{2\pi^2D_{cl}}$ independent of $M$, 
which is confirmed numerically \cite{matsuifull}. It again coincides with the 
Heisenberg time.

However, in the general case of $\omega\neq 0$ the lifetime is enhanced much more 
than the case of $\omega=0$. If $\omega\neq 0$ the statistical distribution of 
$|\delta_{mn}^\pm|=|\gamma_m-\gamma_n\pm \omega T|$ does not suffer from any 
restriction like level repulsion in the vicinity of $|\delta_{mn}^\pm|\sim 0$. 
Hence, unlike the case of $\omega=0$, $|\delta_{mn}^\pm|$ can be arbitrarily small, 
which will make the term $1/\sin^2(\delta_{mn}^{\pm}/2)$ larger. As $M$ increases 
the chance of encountering smaller $|\delta_{mn}^\pm|$ increases.  In the summation over 
$n$ in the RHS of Eq.(\ref{satu}) the term with the smallest  $\delta_{mn}^\pm$
is most dominant. We take here a hypothesis that $|\delta_{mn}^\pm|~(n=1,2...N)$ are 
uniformly distributed independent stochastic variables in the range $[0,2\pi]$. 
This is a rather bold hypothesis that the distribution of chaotic eigenangle is 
under no restriction except for the repulsion of the nearest neighbouring levels, 
but this is not correct in a strict sense.  
It is now easy to show that the probability of
the minimal $\{|\delta_{mn}^\pm|\}_n$ takes a value $x_m$ is given by
$p(x_m) = N/2\pi \e^{-\frac{N}{2\pi}x_m}$. Next, we have to take the sum over $m$. 
At this second stage it is quite plausible to suppose that $x_m$ is now 
statistically independent variable, then it is shown that the probability of 
$\min\{x_1,....x_M\}$ takes a value $x$ is $P(x)=\frac{MN}{2\pi}\e^{-\frac{MN}{2\pi}x}$, 
and the average of the minimal $|\delta_{mn}^\pm/2|$ is $\pi/MN$. 
Then the most dominant term of $1/\sin^2(\delta_{mn}^\pm/2)$ in 
the RHS is $(MN/\pi)^2$. Since $|\<m|\v|n\>|^2 \sim \<|v|^2\> = \corr(0)/N$ 
and $|C_m|^2\sim 1/M$ we may evaluate  $J_\infty^2=D_0\corr(0)MN/2\pi^2$. 
The $\tlife$ is decided by $D_{cl}\tlife=J_\infty^2$ to yield
\begin{eqnarray}
\label{tlife}
        \tlife  =  \frac{D_0\corr(0)MN}{2\pi^2D_{cl}}
\end{eqnarray}
This is our final result and it agrees well 
with the numerical results except for a numerical
factor $\sim 3$ as depicted in Fig.\ref{Fig3}(b).
We have confirmed all the results discussed above are valid also for the coupled 
quantum standard maps. We can expect that it is the general features of ideally 
chaotic quantum maps.

In conclusion, we proposed a method to observe the irreversibility potential 
in quantum system and the associated quantum states. Applying it to coupled KRs,
the relation of the organization of irreversibility to the development
of the entanglement between the constituent systems has been elucidated:
the lifetime of irreversibility largely fluctuates in correlation with the 
anomalous fluctuation of entanglement at the birth of irreversibility, and
it further grows up to proportional to square of Hilbert space dimension in the 
full entanglement regime. 
We finally comment that the linear oscillator L can 
be replaced by a two level system without any essential modifications 
which opens the possibility of experimental implementation on the 
optical lattice setting \cite{matsuifull}.

This work is partly supported by Japanese people's tax via JPSJ KAKENHI 15H03701,
and the authors would like to acknowledge them.
They are also very grateful to Kankikai, Dr.S.Tsuji, and  Koike memorial
house for using the facilities during this study.

\end{document}